\documentclass[a4paper,11pt]{article}
\usepackage{pos}

\newcommand{\Pom}{\mathbb{P}}
\newcommand{\Ode}{\mathbb{O}}

\renewcommand\slash[1]{\not \! #1}

\newcommand{\bpta}{\mbox{\boldmath $p_{t,1}$}}
\newcommand{\bptb}{\mbox{\boldmath $p_{t,2}$}}

\title{Searching for odderon exchange in exclusive $pp \to pp \phi$ and $pp \to pp \phi\phi$ reactions at the LHC}
\ShortTitle{Searching for odderon exchange in exclusive reactions at the LHC}

\author*[a]{P. Lebiedowicz}
\author[b]{O. Nachtmann}
\author[a,\dag]{A. Szczurek}
\notes{\note{Also at \textit{College of Natural Sciences, 
Institute of Physics, University of Rzesz{\'o}w, 
Pigonia 1, PL-35310 Rzesz{\'o}w, Poland}.}}
\affiliation[a]{Institute of Nuclear Physics Polish Academy of Sciences,\\ 
Radzikowskiego 152, PL-31342 Krak{\'o}w, Poland}

\affiliation[b]{Institut f\"ur Theoretische Physik, Universit\"at Heidelberg,\\
Philosophenweg 16, D-69120 Heidelberg, Germany}

\emailAdd{Piotr.Lebiedowicz@ifj.edu.pl}
\emailAdd{O.Nachtmann@thphys.uni-heidelberg.de}
\emailAdd{Antoni.Szczurek@ifj.edu.pl}

\abstract{The possibility to use the exclusive 
$pp \to pp \phi$ and $pp \to pp \phi \phi$ reactions 
in identifying the odderon exchange is discussed. 
So far there is no unambiguous experimental evidence 
for the odderon, the charge conjugation $C = -1$ 
counterpart of the $C = +1$ pomeron. 
Recently proposed tensor-pomeron and vector-odderon model 
for soft high-energy reactions is applied.
For the $p p \to p p \phi$ reaction at high energies
the photon-pomeron fusion is the dominant process 
and the odderon-pomeron fusion is an interesting alternative.
Adding odderon exchange 
improves considerably description of the proton-proton angular 
correlations measured by the WA102 collaboration.
The $p p \to p p \phi \phi$ process 
via pomeron-pomeron fusion is advantageous one 
as here the odderon does not couple to protons.
The observation of large $M_{\phi \phi}$ and 
the rapidity difference $Y_{\phi \phi}$ seems 
well suited to identify odderon exchange.
Comparisons with data from the WA102 experiment are made
and predictions for LHC experiments are given.}

\FullConference{%
  40th International Conference on High Energy physics - ICHEP2020\\
  July 28 - August 6, 2020\\
  Prague, Czech Republic (virtual meeting)
}

\begin{document}
\maketitle

%~~~~~~~~~~~~~~~~~~~~~~~~~~~~~~~~~~~~~~~~~~~~~~~~~~~~~~~~
\section{Introduction}
%~~~~~~~~~~~~~~~~~~~~~~~~~~~~~~~~~~~~~~~~~~~~~~~~~~~~~~~~
\vspace{-0.3cm}

The odderon was introduced on theoretical 
grounds in \cite{Lukaszuk:1973nt}.
It was predicted in QCD as a colourless charge-conjugation 
$C$-odd three-gluon bound state exchange 
\cite{Kwiecinski:1980wb,Bartels:1980pe}.
Recent experimental results by the TOTEM Collaboration 
\cite{Antchev:2017yns, Antchev:2018rec}
have brought the odderon question 
in proton-proton elastic scattering
to the forefront again.
It is of great importance to study
possible odderon effects 
in other reactions.
As was discussed in \cite{Schafer:1991na} 
exclusive diffractive $J/\psi$ and $\phi$ production 
from the pomeron-odderon fusion
in high-energy $pp$ and $p\bar{p}$ collisions 
is a direct probe for a possible odderon exchange.
We shall argue here that the central exclusive production
(CEP) of a $\phi\phi$ state offers
a very nice way to look for odderon effects 
as suggested in \cite{Ewerz:2003xi}.

\begin{figure}[!ht]
\begin{center}
(a)\includegraphics[width=0.3\textwidth]{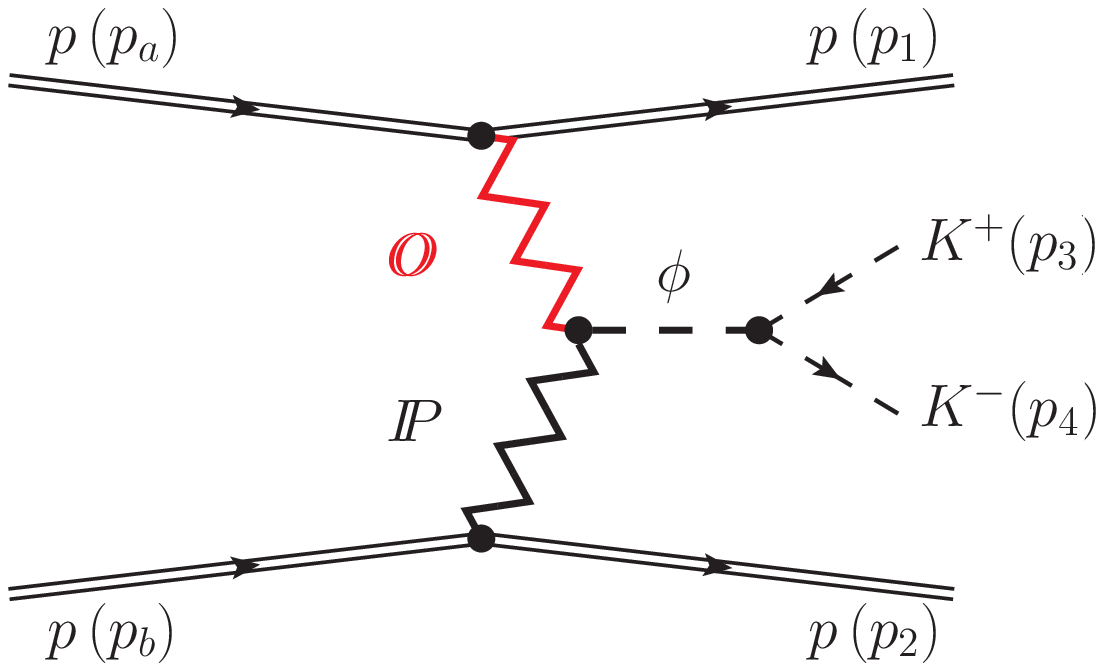} 
(b)\includegraphics[width=0.29\textwidth]{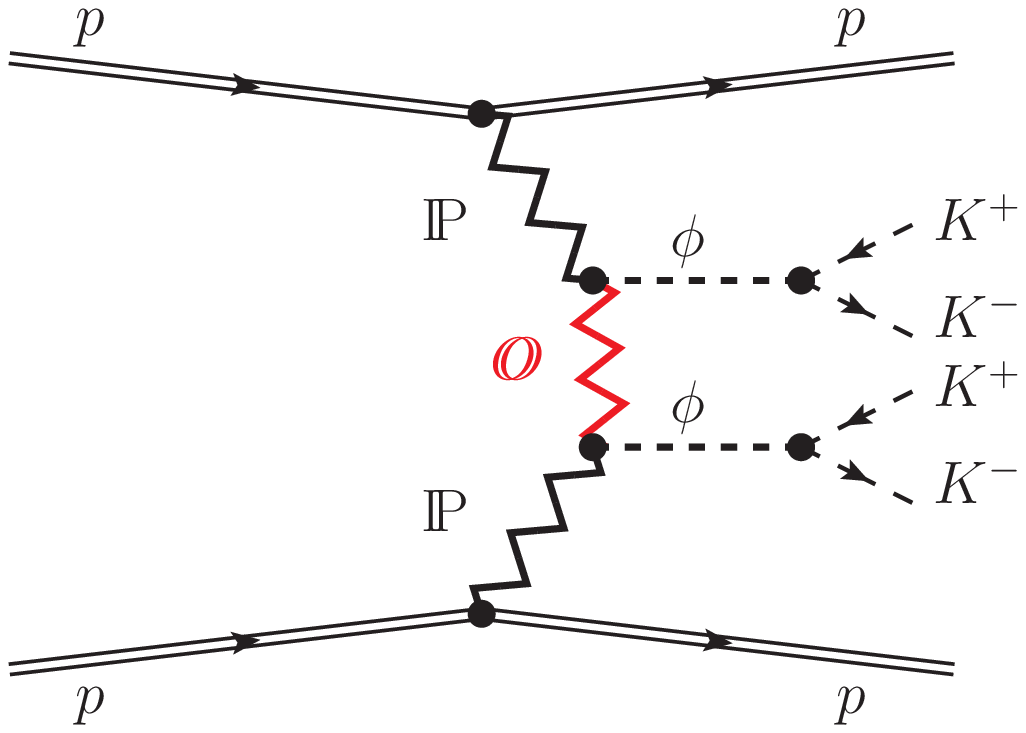} 
\caption{
Diagrams for (a) single $\phi$ production
and for (b) double $\phi$ production
with odderon ($\Ode$) exchange.
There are also diagrams
with the r{\^o}le of the protons interchanged, 
$(p\, (p_{a}), p\, (p_{1})) \leftrightarrow 
 (p\, (p_{b}), p\, (p_{2}))$.}
\label{fig:diagrams}
\end{center}
\end{figure}

In this contribution we will be concerned with CEP 
of single and double $\phi(1020)$ meson production 
observed in the $K^+ K^-$ or $\mu^{+}\mu^{-}$ channels
in $pp$ collisions 
as a possible source of information for soft odderon exchange
(see figure~\ref{fig:diagrams}).
The presentation is based on \cite{Lebiedowicz:2019boz} 
where all details and many more results can be found.
At high energies the $pp \to pp \phi$ reaction 
should be mainly due to photon-pomeron exchange.
The odderon-pomeron fusion mechanism is 
an interesting alternative.
The $pp \to pp \phi\phi$ reaction should be mainly
due to double-pomeron exchange
with resonant production at low $\phi\phi$ invariant masses
and the continuum processes 
(reggeized-$\phi$-meson and odderon exchanges) 
at higher $M_{\phi \phi}$.
The process with an intermediate 
$\hat{t}/\hat{u}$-channel odderon exchange
($\Pom$-$\Ode$-$\Pom$) is a good candidate 
for the odderon searches, as it does not involve 
the coupling of the odderon to the proton.

%--------------------------------------------------------
%--------------------------------------------------------

\vspace{-0.8cm}
We treat our reactions 
in the tensor-pomeron and vector-odderon approach
as introduced in~\cite{Ewerz:2013kda}.
This approach has a good basis from nonperturbative QCD
using functional integral techniques~\cite{Nachtmann:1991ua}.
We describe the pomeron and the $C = +1$
reggeons as effective rank 2 symmetric tensor exchanges,
the odderon and $C = -1$ reggeons as effective vector exchanges.
There are by now many applications of the tensor-pomeron model
to two-body hadronic reactions \cite{Ewerz:2016onn},
to photoproduction,
to DIS structure functions at low $x$, 
and especially to CEP reactions:
$p + p \to p + X + p$, where
$X = \eta,\, \eta',\, f_{0},\, f_{1},\, f_{2},\, \pi^{+}\pi^{-},\,
p \bar{p},\, K\bar{K}, \, 4\pi,\,  4K,\, \rho^{0},\, \phi,\, \phi \phi$;
see e.g. \cite{Lebiedowicz:2013ika,Lebiedowicz:2016ioh,
Lebiedowicz:2019boz,Lebiedowicz:2019jru,Lebiedowicz:2020yre}.

\vspace{-0.3cm}
%~~~~~~~~~~~~~~~~~~~~~~~~~~~~~~~~~~~~~~~~~~~~~~~~~~~~~~~~
\section{A sketch of formalism}
%~~~~~~~~~~~~~~~~~~~~~~~~~~~~~~~~~~~~~~~~~~~~~~~~~~~~~~~~
\vspace{-0.3cm}

As an example, we consider the reaction
\begin{equation}
p(p_{a}) + p(p_{b}) \to
p(p_{1}) + 
[ \phi(p_{34}) \to K^{+}(p_{3}) + K^{-}(p_{4}) ]
+ p(p_{2}) \,,
\label{2to4_reaction_KK_via_phi}
\end{equation}
where $p_{a,b}$, $p_{1,2}$ 
denote the four-momenta of the protons and
$p_{3,4}$ denote the four-momenta of the $K$ mesons, respectively.
For high energies and central $\phi$ production we 
expect the reaction (\ref{2to4_reaction_KK_via_phi})
to be dominated by the fusion processes 
$\gamma \Pom \to \phi$ and $\Ode \Pom \to \phi$.
For the first process all couplings are, in essence, known.
The parameters of $\phi$ photoproduction
were fixed to describe the HERA data 
taking into account the $\phi$-$\omega$ mixing effect
\cite{Lebiedowicz:2019boz}.
For the odderon-exchange process we shall use the ans{\"a}tze 
from \cite{Ewerz:2013kda}
and we shall try to get information on the odderon parameters 
and couplings from the comparison to the WA102 data
for the $pp \to pp \phi$ and $pp \to pp \phi \phi$ reactions.
Of course, at the relatively low c.m. energy 
of the WA102 experiment,
$\sqrt{s} = 29.1$~GeV, we have to include also 
subleading contributions with reggeized-vector-meson 
(or reggeon) exchanges discussed
in \cite{Lebiedowicz:2019boz}.

The Born-level amplitude for the diffractive production of the $\phi(1020)$ 
via odderon-pomeron fusion
[figure~\ref{fig:diagrams}(a)]
can be written as
\begin{eqnarray}
&&{\cal M}^{(\Ode \Pom)}_{pp \to pp K^{+}K^{-}} 
= (-i)
\bar{u}(p_{1}, \lambda_{1}) 
i\Gamma^{(\Ode pp)}_{\mu}(p_{1},p_{a}) 
u(p_{a}, \lambda_{a})\,
i\Delta^{(\Ode)\,\mu \rho_{1}}(s_{1}, t_{1}) \,
i\Gamma^{(\Pom \Ode \phi)}_{\rho_{1} \rho_{2} \alpha \beta}(-q_{1},p_{34}) \nonumber \\
&& \qquad \quad \times  \;
i\Delta^{(\phi)\,\rho_{2} \kappa}(p_{34})\,
i\Gamma^{(\phi KK)}_{\kappa}(p_{3},p_{4})\,
i\Delta^{(\Pom)\,\alpha \beta, \delta \eta}(s_{2},t_{2}) \,
\bar{u}(p_{2}, \lambda_{2}) 
i\Gamma^{(\Pom pp)}_{\delta \eta}(p_{2},p_{b}) 
u(p_{b}, \lambda_{b}) \,. \qquad
\label{amplitude_oderon_pomeron}
\end{eqnarray}
The kinematic variables are
$p_{34} = p_{3} + p_{4}$,
$q_1 = p_{a} - p_{1}$,
$q_2 = p_{b} - p_{2}$,
$t_1 = q_{1}^{2}$,
$t_2 = q_{2}^{2}$,
$s = (p_{a} + p_{b})^{2}$,
$s_{1} = (p_{1} + p_{34})^{2}$,
$s_{2} = (p_{2} + p_{34})^{2}$.
The effective pomeron propagator and the pomeron-proton vertex function are as follows \cite{Ewerz:2013kda}:
\begin{eqnarray}
&&i \Delta^{(\Pom)}_{\mu \nu, \kappa \lambda}(s,t) =
\frac{1}{4s} \left( g_{\mu \kappa} g_{\nu \lambda} 
                  + g_{\mu \lambda} g_{\nu \kappa}
                  - \frac{1}{2} g_{\mu \nu} g_{\kappa \lambda} \right)
(-i s \alpha'_{\Pom})^{\alpha_{\Pom}(t)-1}\,,
\label{A1}\\
&&i\Gamma_{\mu \nu}^{(\Pom pp)}(p',p)
=-i 3 \beta_{\Pom NN} F_{1}( (p'-p)^{2} )
\left\lbrace 
\frac{1}{2} 
\left[ \gamma_{\mu}(p'+p)_{\nu} 
     + \gamma_{\nu}(p'+p)_{\mu} \right]
- \frac{1}{4} g_{\mu \nu} ( \slash{p}' + \slash{p} )
\right\rbrace\,, \qquad
\label{A4} 
\end{eqnarray}
where $\beta_{\Pom NN} = 1.87$~GeV$^{-1}$ and $t = (p'-p)^{2}$.
For simplicity we use the electromagnetic Dirac form factor $F_{1}(t)$ of the proton.
The pomeron trajectory $\alpha_{\Pom}(t)$
is assumed to be of standard linear form:
$\alpha_{\Pom}(t) = \alpha_{\Pom}(0)+\alpha'_{\Pom}\,t$,
with
$\alpha_{\Pom}(0) = 1.0808$ and
$\alpha'_{\Pom} = 0.25 \; {\rm GeV}^{-2}$.

Our ansatz for the $C = -1$ odderon follows (3.16), (3.17) and (3.68), (3.69) 
of \cite{Ewerz:2013kda}:
\begin{eqnarray}
&&i \Delta^{(\Ode)}_{\mu \nu}(s,t) = 
-i g_{\mu \nu} \frac{\eta_{\Ode}}{M_{0}^{2}} \,(-i s \alpha'_{\Ode})^{\alpha_{\Ode}(t)-1}\,,
\label{A12} \\
&&i\Gamma_{\mu}^{(\Ode pp)}(p',p) 
= -i 3\beta_{\Ode pp} \,M_{0}\,F_{1} ( (p'-p)^{2} ) \,\gamma_{\mu}\,,
\label{A13}
\end{eqnarray}
where $\eta_{\Ode}$ is a parameter with value 
$\eta_{\Ode} = \pm 1$;
$M_{0} = 1$~GeV is inserted for dimensional reasons.
We assumed 
$\beta_{\Ode pp} = 0.1 \times \beta_{\Pom NN}$.
We take
$\alpha_{\Ode}(t) = \alpha_{\Ode}(0)+\alpha'_{\Ode}\,t$.
In our calculations we shall choose as default values 
$\alpha_{\Ode}(0) = 1.05$,
$\alpha'_{\Ode} = 0.25 \,\mathrm{GeV}^{-2}$, 
and $\eta_{\Ode} = - 1$; see \cite{Lebiedowicz:2019boz}.

For the $\Pom \Ode \phi$ vertex we use an ansatz analogous to the $\Pom \phi \phi$ vertex
(see (3.48)--(3.50) of \cite{Lebiedowicz:2019jru})
\begin{eqnarray}
i\Gamma^{(\Pom \Ode \phi)}_{\rho_{1} \rho_{2} \alpha \beta}(-q_{1},p_{34}) 
&=&
i \left[ 2\,a_{\Pom \Ode \phi}\, \Gamma^{(0)}_{\rho_{2} \rho_{1} \alpha \beta}(p_{34},-q_{1})
- b_{\Pom \Ode \phi}\,\Gamma^{(2)}_{\rho_{2} \rho_{1} \alpha \beta}(p_{34},-q_{1}) \right] \nonumber\\
&&\times F_{M}(q_{2}^{2})\,F_{M}(q_{1}^{2})\,F^{(\phi)}(p_{34}^{2}) \,.
\label{A15}
\end{eqnarray}  
Here we use the relations (3.20) of \cite{Ewerz:2013kda}
and as in (3.49) of \cite{Lebiedowicz:2019jru}
we take the factorised form for the $\Pom \Ode \phi$ form factor; see \cite{Lebiedowicz:2019boz}.
The coupling parameters $a_{\Pom \Ode \phi}$, $b_{\Pom \Ode \phi}$ in (\ref{A15})
and the cut-off parameter $\Lambda_{0,\,\Pom \Ode \phi}^{2}$ 
in $F_{M}(t) = 1/(1 - t/\Lambda_{0,\,\Pom \Ode \phi}^{2})$
could be adjusted to experimental data.
The WA102 data allow us to determine the respective
coupling constants as $a_{\Pom \Ode \phi} = -0.8$~GeV$^{-3}$, $b_{\Pom \Ode \phi}= 1.6$~GeV$^{-1}$, and
$\Lambda_{0,\,\Pom \Ode \phi}^{2} = 0.5$~GeV$^{2}$\cite{Lebiedowicz:2019boz}.
We have checked that these parameters 
are compatible with our analysis of the WA102 data for
the $pp \to pp \phi \phi$ reaction in \cite{Lebiedowicz:2019jru}.

The full amplitude includes the $pp$-rescattering corrections 
in the eikonal approximation; see~\cite{Lebiedowicz:2019boz}.

%~~~~~~~~~~~~~~~~~~~~~~~~~~~~~~~~~~~~~~~~~~~~~~~~~~~~~~~~
\section{Results}
%~~~~~~~~~~~~~~~~~~~~~~~~~~~~~~~~~~~~~~~~~~~~~~~~~~~~~~~~
\vspace{-0.3cm}

\begin{figure}[!ht]
\begin{center}
\includegraphics[width=0.43\textwidth]{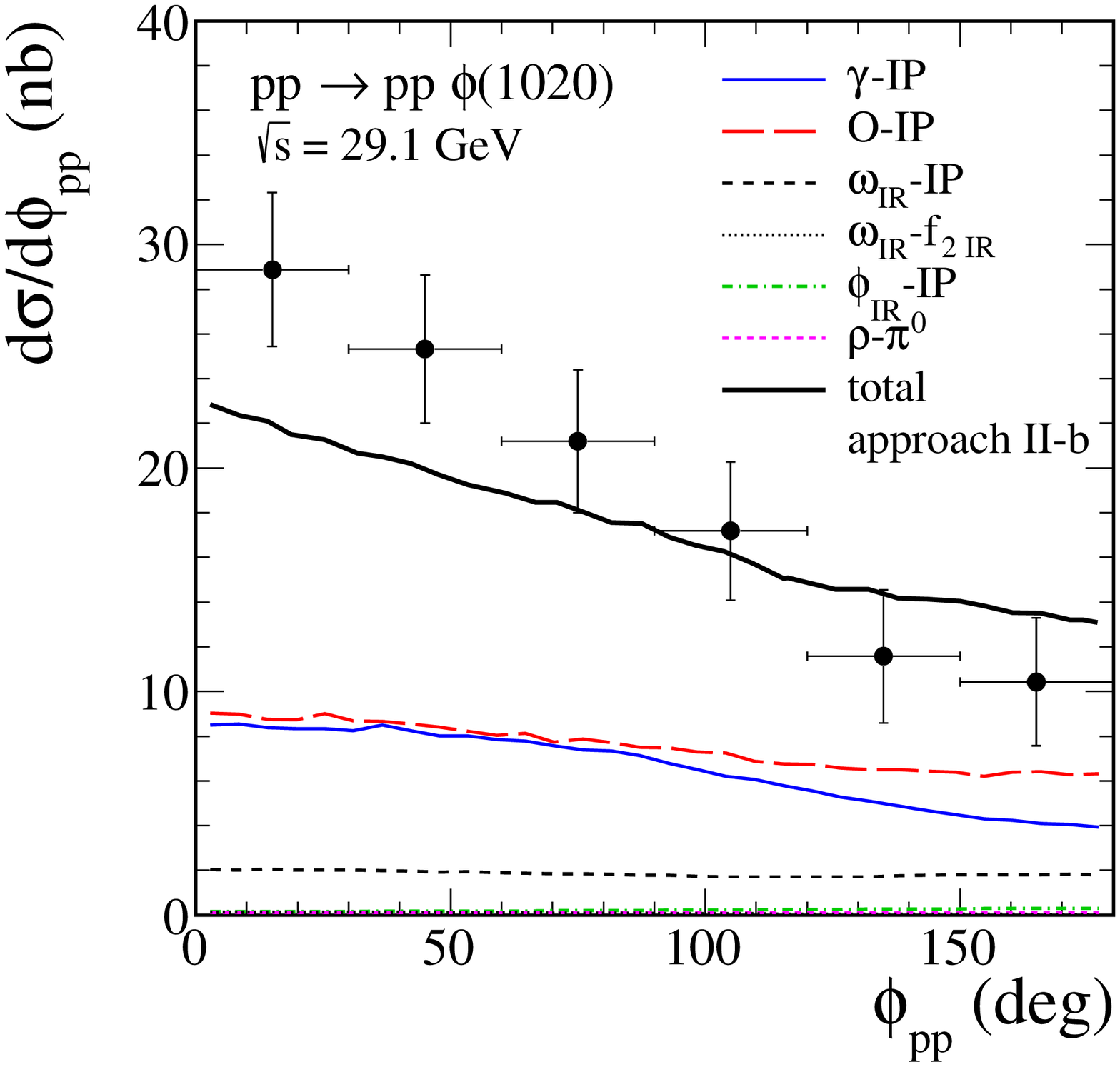}
\includegraphics[width=0.43\textwidth]{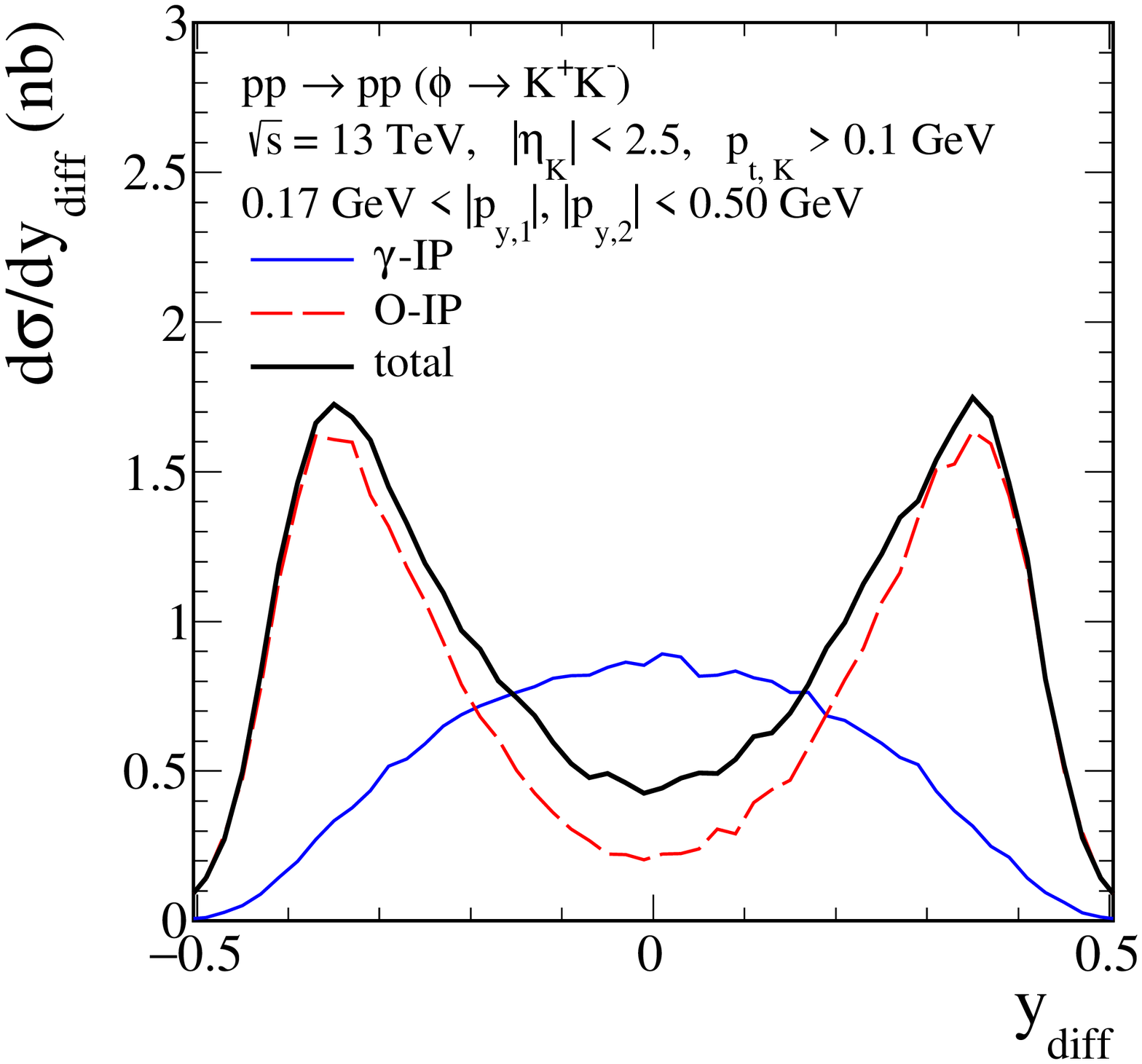}
\caption{\label{fig:LHC_KK}
Left panel: The distributions in $\phi_{pp}$ 
together with
the WA102 experimental data points for $\sqrt{s} = 29.1$~GeV 
normalized to the central value 
of the total cross section $\sigma_{\rm exp} = 60$~nb 
from \cite{Kirk:2000ws}.
The coherent sum of all terms is shown by the black solid line.
Right panel: The distribution in rapidity difference between kaons 
for the $pp \to pp (\phi \to K^{+}K^{-})$ reaction 
for the ATLAS-ALFA kinematics.}
\end{center}
\end{figure}

It is very difficult to describe the WA102 data from 
\cite{Kirk:2000ws}
for the $pp \to pp \phi$ reaction including the
$\gamma \Pom$-fusion mechanism only.
As was presented in \cite{Lebiedowicz:2019boz}
inclusion of the odderon-exchange contribution significantly
improves the description of the $pp$ azimuthal correlations 
($\phi_{pp}$ is angle between the transverse momentum vectors 
$\bpta$, $\bptb$ of the outgoing protons)
and the ${\rm dP_{t}} = |\bptb - \bpta|$ dependence 
of $\phi$ CEP measured by the WA102 collaboration.
The absorption effects - very important - were included in the calculations. 
In the left panel of figure~\ref{fig:LHC_KK}
we present the $\Ode$-$\Pom$ contribution
(approach~II of \cite{Lebiedowicz:2019boz})
together with the $\gamma$-$\Pom$ contribution
and with the subleading terms. 
Adding odderon exchange term
improves description of the proton-proton angular 
correlations.
Having fixed the parameters of our model 
to the WA102 data we show our predictions 
at $\sqrt{s} = 13$~TeV for the LHC.
Here we focus on the limited invariant mass region
around the $\phi(1020)$ resonance.

%--------------------------------------------------------
%--------------------------------------------------------

%--------------------------------------------------------
\begin{figure}[!ht]
\begin{center}
\includegraphics[width=0.43\textwidth]{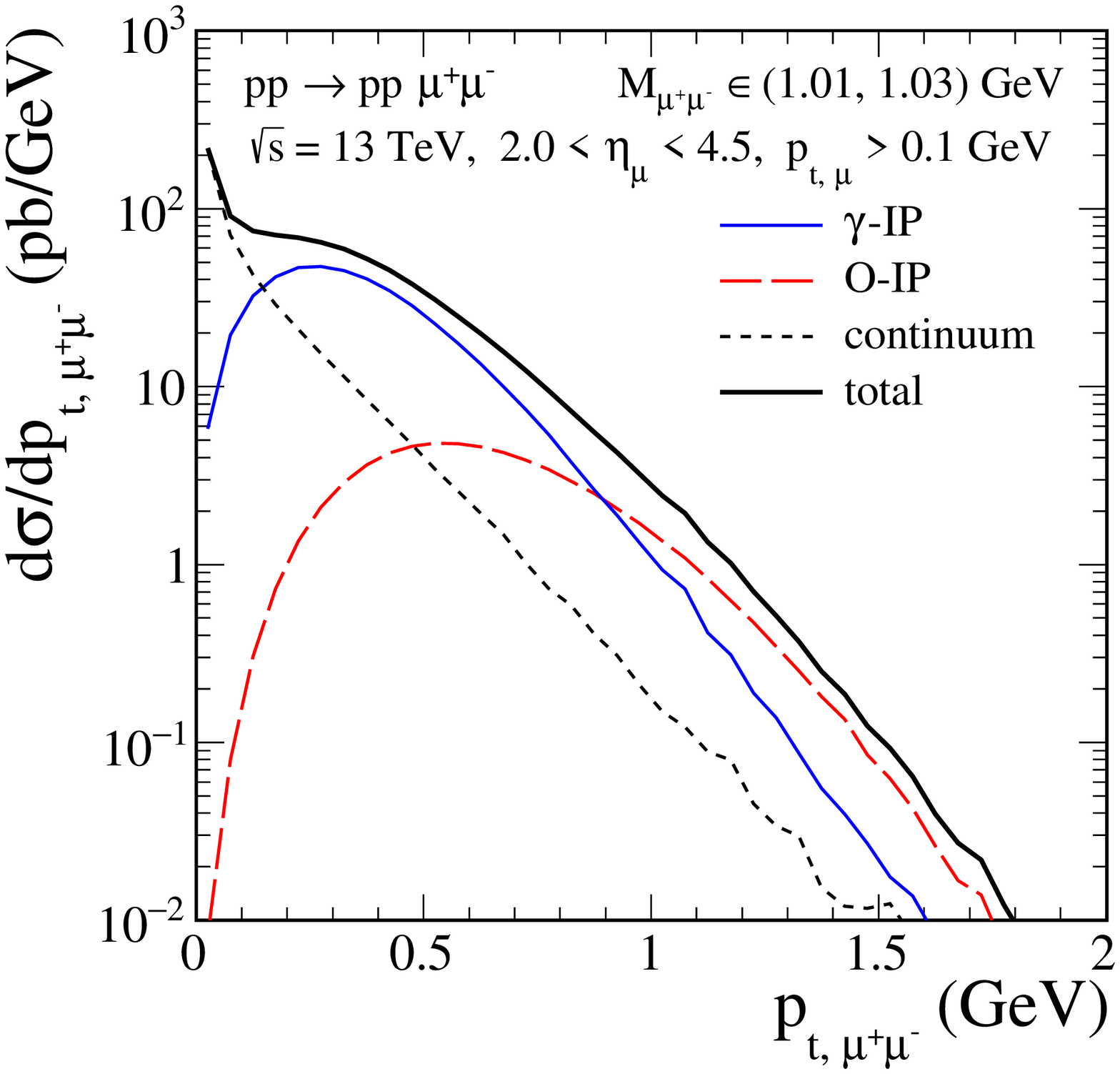}
\includegraphics[width=0.43\textwidth]{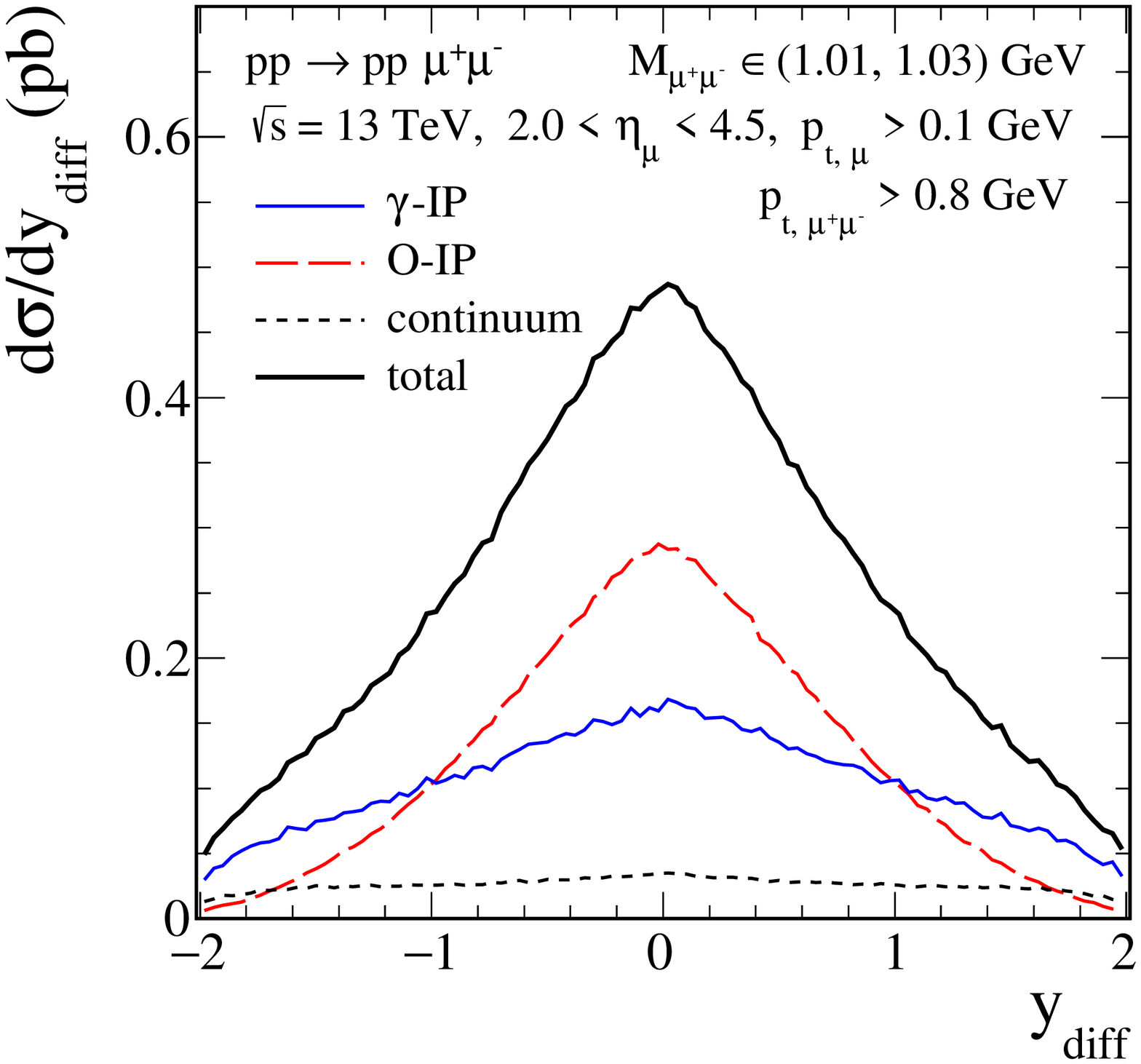}
\caption{\label{fig:LHCb_mumu_3}
The distributions in transverse momentum of the $\mu^{+}\mu^{-}$ 
pair (left)
and in rapidity difference between muons (right)
for the $pp \to pp \mu^{+}\mu^{-}$ reaction for $\sqrt{s} = 13$~TeV
and $M_{\mu^{+}\mu^{-}} \in (1.01, 1.03)$~GeV.
Results for the $\gamma$-$\Pom$ and $\Ode$-$\Pom$ fusion terms,
the continuum term as well as their coherent sum are shown.}
\end{center}
\end{figure}
%--------------------------------------------------------
\vspace{-0.6cm}
In the right panel of figure~\ref{fig:LHC_KK}
we show the results for the $pp \to pp (\phi \to K^{+}K^{-})$
reaction for experimental conditions relevant for ATLAS-ALFA or CMS-TOTEM.
The $\Ode$-$\Pom$ contribution dominates at larger
$p_{t, K^{+}K^{-}}$ (or transverse momentum of the $K^+ K^-$ pair) and
$|\rm{y_{diff}}|$ compared to the $\gamma$-$\Pom$ contribution.
For the ATLAS-ALFA kinematics 
the absorption effects lead to a large
damping of the cross sections 
for both the mechanisms;
see Table~II of \cite{Lebiedowicz:2019boz}.

Now we discuss the $pp \to pp \mu^{+}\mu^{-}$ reaction
at forward rapidities and without measurement of protons
relevant for LHCb.
Figure~\ref{fig:LHCb_mumu_3} shows the distribution
in transverse momentum of the $\mu^{+}\mu^{-}$ pair.
We can see that the low-$p_{t,\mu^{+}\mu^{-}}$ cut can be helpful
to reduce the dimuon-continuum
and $\gamma$-$\Pom$-fusion contributions.
In the right panel we show the $\rm{y_{diff}}$ (rapidity difference between muons) distribution
when imposing in addition a cut $p_{t, \mu^{+}\mu^{-}} > 0.8$~GeV.
The $\gamma \gamma \to \mu^{+}\mu^{-}$ continuum contribution 
is now very small.
At $\rm{y_{diff}} = 0$ the $\Ode$-$\Pom$ term
should win with the $\gamma$-$\Pom$ term.
In contrast to dikaon CEP here there is 
for both contributions
a maximum at $\rm{y_{diff}} = 0$.

\begin{figure}[!ht]
\begin{center}
\includegraphics[width=0.43\textwidth]{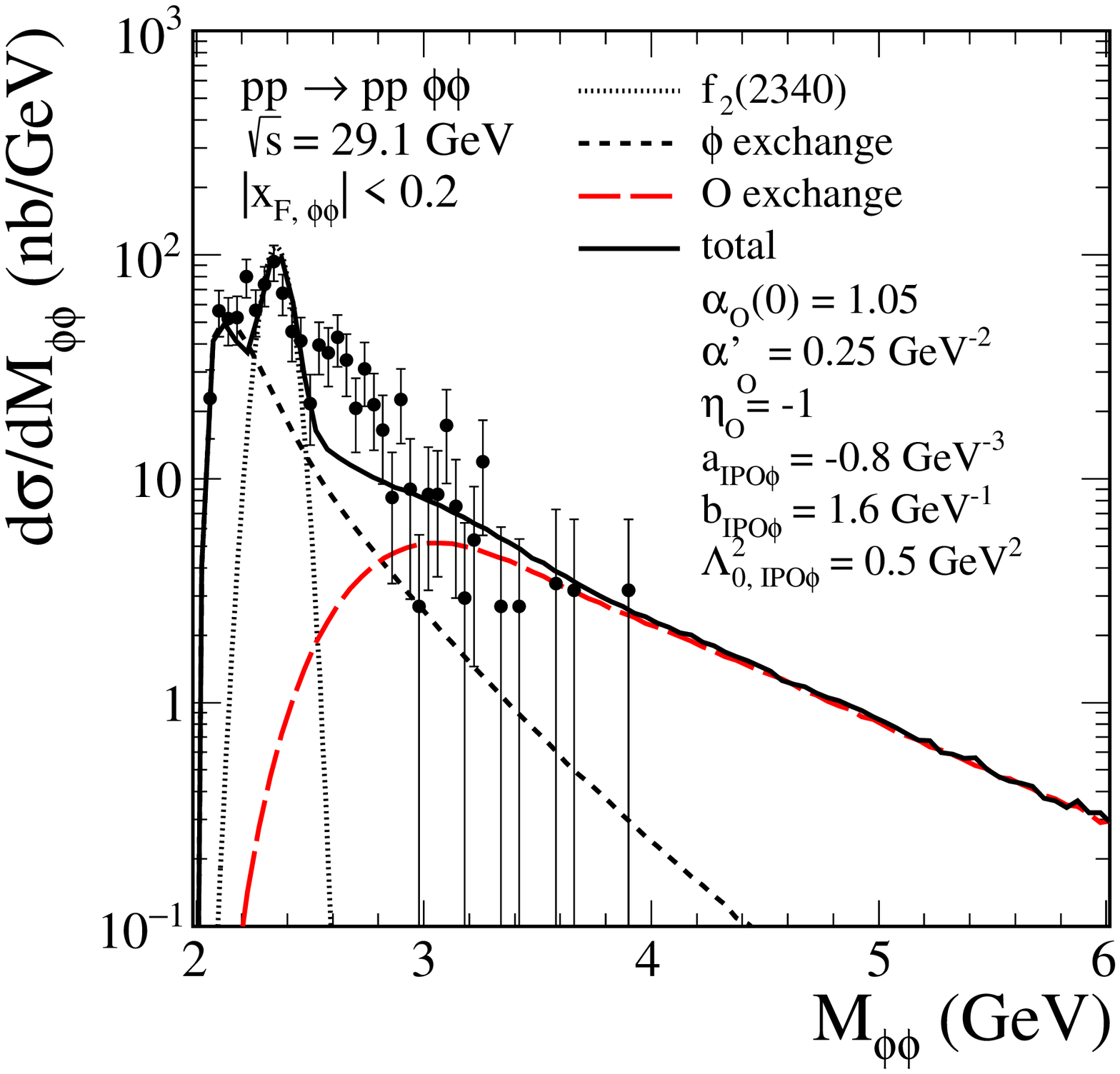}
\includegraphics[width=0.43\textwidth]{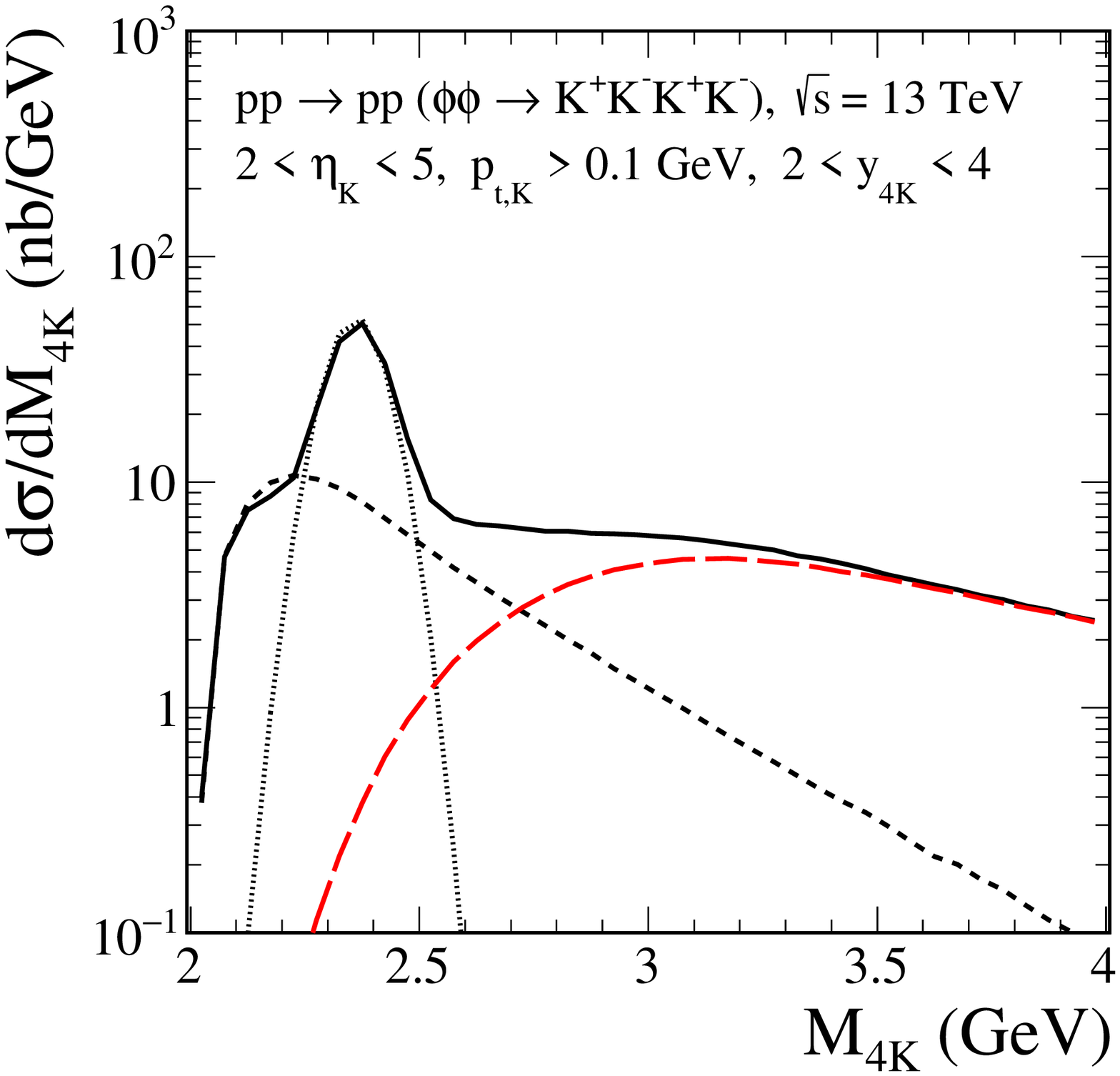}
\caption{The distributions in $\phi\phi$ invariant mass 
(left) for $\sqrt{s} = 29.1$~GeV together with 
the WA102 data from \cite{Barberis:1998bq}
and (right) in ${\rm M}_{4K}$ for the LHCb kinematics.
The short-dashed line corresponds to 
the reggeized-$\phi$-exchange contribution,
the dotted line corresponds to the $f_{2}(2340)$ contribution,
the red long-dashed line represents the $\Ode$-exchange contribution.
The coherent sum of all terms is shown by the black solid line.}
\label{fig:odderon_LHC}
\end{center}
\end{figure}

Now we go to the $pp \to pp \phi \phi$ reaction.
Figure~\ref{fig:odderon_LHC} shows the results
including the $f_{2}(2340)$ term
and the continuum processes due to reggeized-$\phi$ 
and odderon exchanges.
For the details how to calculate these processes see \cite{Lebiedowicz:2019jru}.
Inclusion of the odderon exchange improves the description of the WA102 data \cite{Barberis:1998bq} for the $pp \to pp \phi \phi$ reaction;
see the left panel of figure~\ref{fig:odderon_LHC}.
Here we showed results for the odderon-exchange
contribution
with the parameters of our model fixed to the WA102 data 
\cite{Kirk:2000ws} on single CEP of $\phi$; 
see section~IV~A of \cite{Lebiedowicz:2019boz}.
In the right panel we show the distribution in four-kaon 
invariant mass for the LHCb experimental conditions.
The small intercept of the $\phi$-reggeon exchange, 
$\alpha_{\phi}(0) = 0.1$
makes the $\phi$-exchange contribution steeply falling 
with increasing ${\rm M}_{4K}$.
Therefore, an odderon with an intercept 
$\alpha_{\Ode}(0)$ around 1.0
should be clearly visible in the region 
of large ${\rm M}_{4K}$ 
(and also for large rapidity distance between the $\phi$ mesons).

%--------------------------------------------------------

%~~~~~~~~~~~~~~~~~~~~~~~~~~~~~~~~~~~~~~~~~~~~~~~~~~~~~~~~
\section{Conclusions}
%~~~~~~~~~~~~~~~~~~~~~~~~~~~~~~~~~~~~~~~~~~~~~~~~~~~~~~~~
\vspace{-0.3cm}

We have discussed in detail 
the $p p \to p p \phi$ and $p p \to p p \phi \phi$ reactions.
For single $\phi$ CEP at the LHC there are two basic processes: 
the relatively well known $\gamma$-$\Pom$ fusion
and the rather elusive $\Ode$-$\Pom$ fusion.
We fixed the parameters of the pomeron-odderon contribution
to obtain a good description of the WA102 data 
\cite{Barberis:1998bq,Kirk:2000ws}.
Then we have estimated the integrated cross sections 
and several differential distributions 
at the LHC; see Table~II of \cite{Lebiedowicz:2019boz}.
It is a main result of our analysis that,
the $\rm{y_{diff}}$ distributions are very different 
for the $\gamma$-$\Pom$- and $\Ode$-$\Pom$-fusion processes. 
The $\mu^+ \mu^-$ channel seems to be less promising
in identifying the odderon exchange 
at least when only the $p_{t, \mu}$ cuts are imposed.
To observe a sizeable deviation from photoproduction
a $p_{t,\mu^+ \mu^-} > 0.8$~GeV cut on the transverse momentum 
of the $\mu^+ \mu^-$ pair seems necessary. 
A combined analysis of 
both the $K^+ K^-$ and the $\mu^+ \mu^-$ channels 
should be the ultimate goal in searches for odderon exchange.
%In our opinion several distributions should be studied 
%to draw a definite conclusion.

The $pp \to pp \phi \phi$ process via odderon exchange
[figure~\ref{fig:diagrams}(b)]
seems promising as here the odderon does not couple to protons.
We find from our model that the odderon-exchange contribution
should be distinguishable from other contributions
for relatively large four-kaon invariant masses
(outside of the region of resonances)
and for large rapidity distance between the $\phi$ mesons.
Hence, to study this type of mechanism one should 
investigate ``three-gap events''
(proton--gap--$\phi$--gap--$\phi$--gap--proton).
Experimentally, this should be a clear signature.

We are looking forward to first experimental 
results on single and double $\phi$ CEP at the LHC.

\vspace{-0.3cm}
\acknowledgments
\vspace{-0.2cm}

The authors thank the organisers of the ICHEP 2020
conference for making this presentation of our results possible.
This work was partially supported by
the NCN Grant No. 2018/31/B/ST2/03537.

\end{document}